\documentclass[slac_one]{revtex4}
\usepackage{graphicx}
\usepackage{fancyhdr}
\usepackage{psfrag}

\psfrag {LD2} {$\bm \ld$}

\newcommand{\ttbar}{{t\bar{t}}}

\begin{document}

\title{Top Quark Pair Production Cross Section and Forward-Backward Asymmetry at the Tevatron} 

\author{A. Lister (on behalf of the CDF and D\O\ Collaborations)}
\affiliation{University of California at Davis, CA 95616, USA}

\begin{abstract}
We present recent results on top quark pair production cross section in several final states and the $\ttbar$ forward-backward charge asymmetry in the lepton + jets channel at the Tevatron $p\bar{p}$ collider operating at $\sqrt{s} = 1.96$ TeV. A new combination of D\O\ measurements gives a $\ttbar$ production cross section of $\sigma_\ttbar$ = 7.83 $^{+ 0.46}_{ - 0.45}$ (stat) $^{+ 0.64}_{ - 0.53}$ (syst) $\pm$ 0.48 (lumi). The new combination of CDF results for $\ttbar$ production yields 7.0 $\pm$ 0.3 (stat) $\pm$ 0.4 (syst) $\pm$ 0.4 (lumi) pb, which corresponds to a total uncertainty of 9\%, very close to the that of the current best theoretical predictions. It is important to measure the $\ttbar$ cross section in as many different channels as possible as any significant discrepancy between them could be a sign of new physics. Three new measurements of the forward-backward charge asymmetry are also presented. The two CDF measurements correct the acceptance and observed asymmetry back to parton level in order to compare the central values with specific theoretical predictions. The D\O\ measurement does not unfold its acceptance and therefore does not depend on the specific method used for unfolding.

\end{abstract}

\maketitle

\thispagestyle{fancy}


\section{INTRODUCTION}
The study of top quark pair production at the Tevatron is now entering the realm of precision physics. With 2.8 fb$^{-1}$ of CDF data there are over 1200 reconstructed $\ttbar$ events in the lepton + jets channel. As the integrated luminosity increases, it is possible to improve not only measurements of the properties of the top quark but also maybe to search for new physics by comparing the $\ttbar$ cross section measurements in different channels, or by measuring the forward-backward charge asymmetry.

We present here results of the $\ttbar$ cross section in the dilepton channel, where both of the $W$s decay leptonically as well as in the lepton + jets channel, where one $W$ decays leptonically and the other hadronically. We will also present combinations of recent CDF and D\O\ $\ttbar$ cross sections. In addition, we present results from studies of asymmetry in the centre of mass in the production of top quarks in the lepton + jets $\ttbar$ final states.

\section{TOP PAIR PRODUCTION CROSS SECTION}
\begin{figure}
  \begin{center}
     \includegraphics[height=70mm]{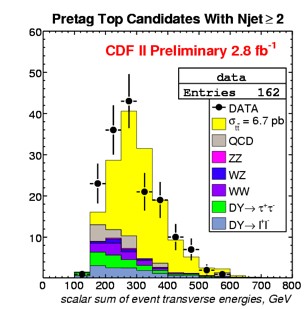}
      \includegraphics[height=70mm]{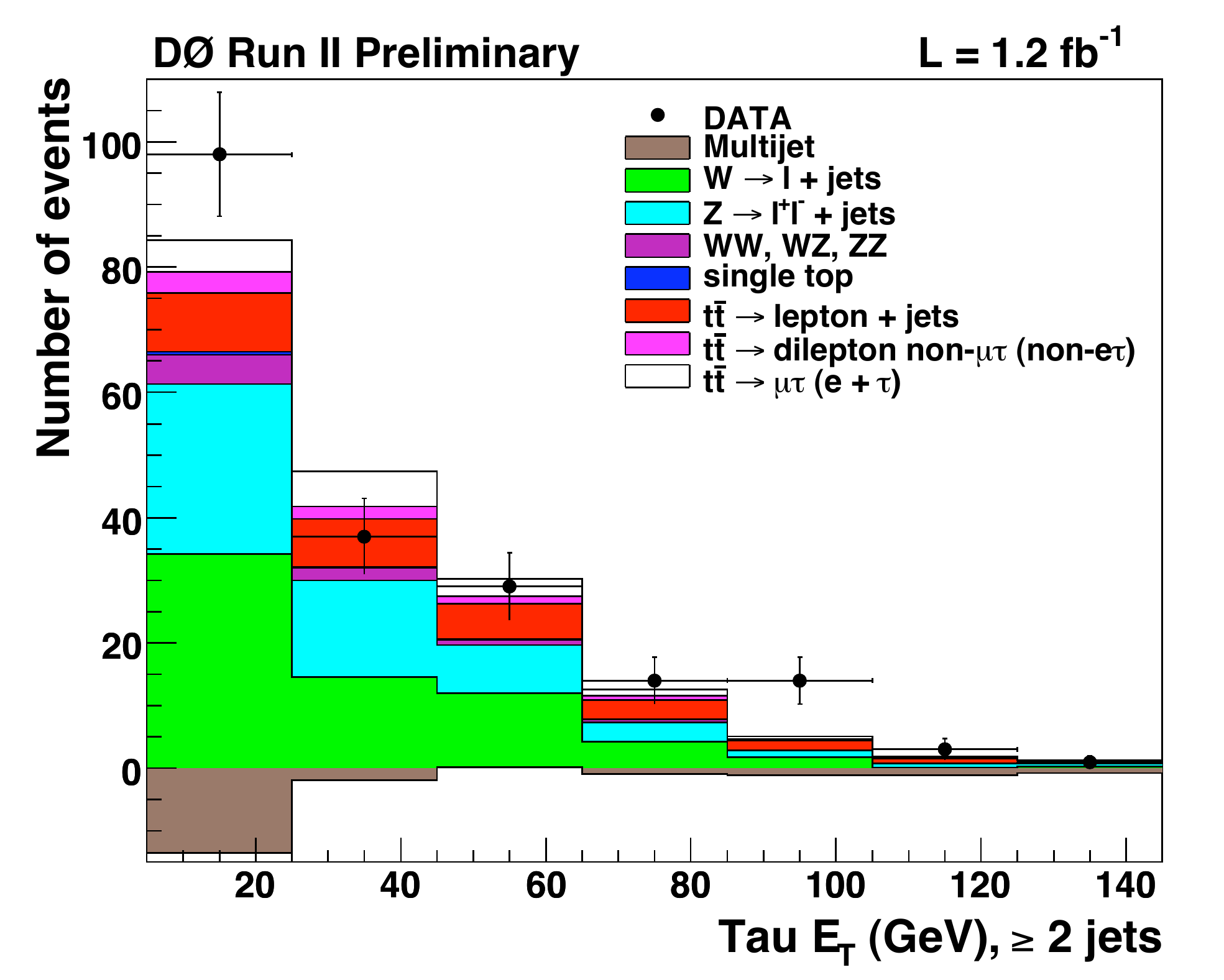}
  \end{center}
  \caption{(left) Scalar sum of the transverse energies in the dilepton data compared to the sum of all the backgrounds and $\ttbar$ signal (yellow) in the $\geq$2 jets channel. (right) $E_T$ distribution of the $\tau$ leptons in data compared to the sum of all backgrounds and $\ttbar$ signal (white).}
  \label{fig:topXs_dil}
\end{figure}

Many new $\ttbar$ cross section measurements were presented at this conference. In the dilepton channel, a new measurement from CDF~\cite{topXs_dil_cdf} uses the full $2.8 fb^{-1}$ of data. In this channel, the backgrounds are relatively small, as illustrated in Fig.~\ref{fig:topXs_dil} (left). The measured cross section is $\sigma_\ttbar$ = 6.7 $\pm$ 0.8 (stat) $\pm$ 0.4 (syst) $\pm$ 0.4 (lumi) pb.

D\O\ has developed excellent hadronic $\tau$ identification algorithms that makes use of a series of Neural Networks to separate out real $\tau$ leptons from electrons and jets. This can be used to measure the $\ttbar$ cross section in the dilepton mode, where one of the $W$s decays to a $\tau$ that subsequently decays to a $\nu_{\tau}$ and hadrons~\cite{topXs_ltau_d0}. The event signature consists of a lepton ($e$ or $\mu$) and a jet from a $\tau$ decay, missing transverse energy and at least two other jets. Although the signal fraction after all selections is relatively small, as can be seen from Fig.~\ref{fig:topXs_dil} (right), it is sufficient for extracting the $\ttbar$ cross section. The result, using 2.2 $fb^{-1}$ of D\O\ data, is $\sigma_\ttbar$ = 7.3 $^{+1.3}_{-1.2}$ (stat) $^{+1.2}_{-1.1}$ (syst) $\pm$ 0.5(lumi) pb, consistent with that of all other channels, thereby showing so large sign of new physics from this sector.

\begin{figure}
  \begin{center}
     \includegraphics[width=88mm]{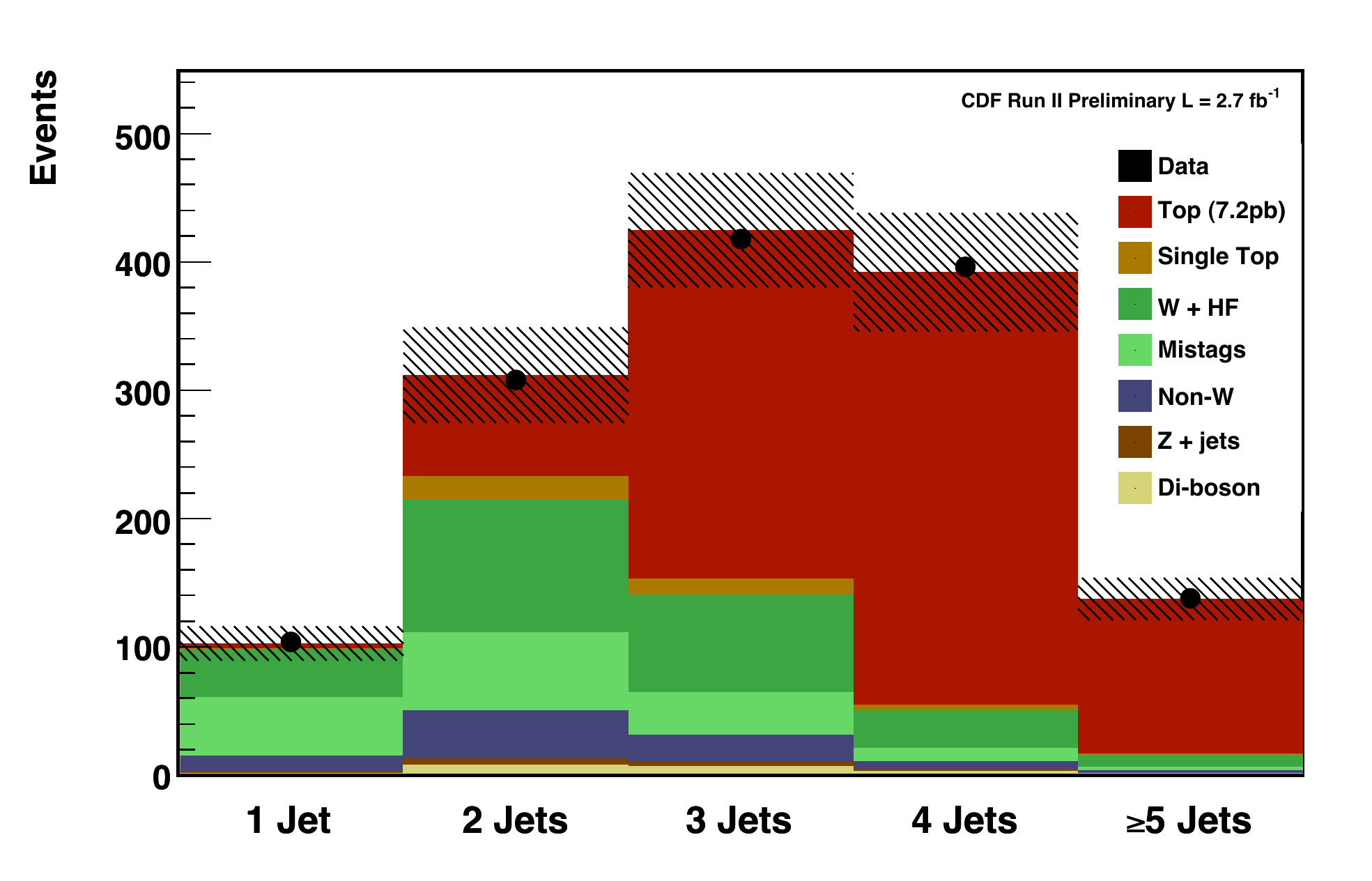}
     \includegraphics[width=88mm]{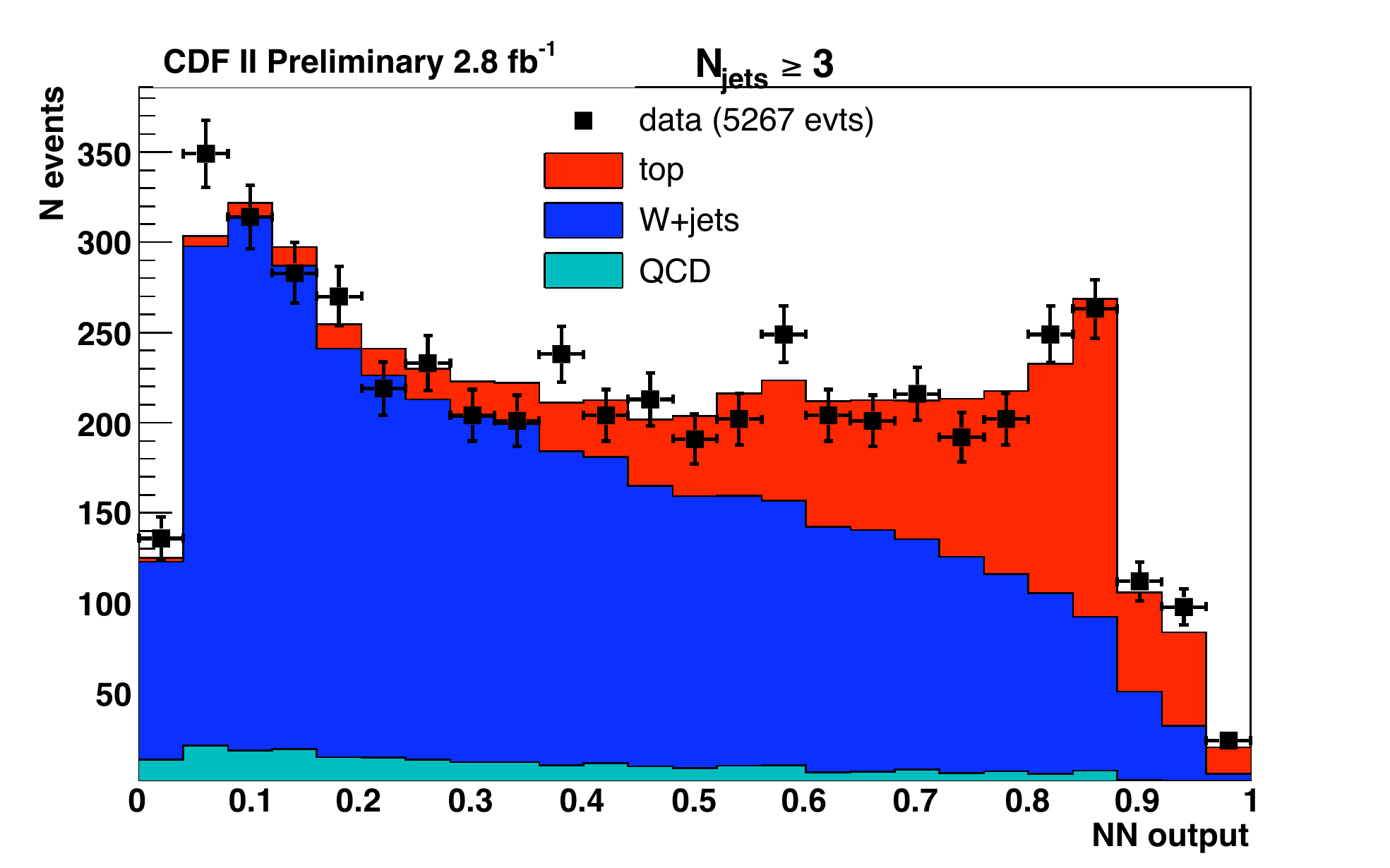}
  \end{center}
  \caption{(left) Distribution of the number of jets in the lepton + jets channel after requiring at least one heavy-flavor tag in each event. The signal $\ttbar$ contribution is shown in red. (right) Neural Network output variable used to fit to the $\ttbar$ cross section in the lepton + jets channel without using any b-tagging requirements. The $\ttbar$ contribution obtained from the fit is shown in red.}
  \label{fig:topXs_lj}
\end{figure}

In the lepton + jets channel, two new CDF measurements were shown that use the full integrated luminosity available, 2.8 $fb^{-1}$. The first makes use of the enhanced signal to background fraction when at least one of the jets is required to have a heavy-flavor tag (using the displaced vertex tagger SecVtx)~\cite{topXs_lj_tag_cdf}. Figure~\ref{fig:topXs_lj} (left) shows the distribution of $\ttbar$ signal and standard model backgrounds as a function of jet multiplicity. In this analysis, the data is used to constrain the backgrounds that cannot be predicted with great accuracy, such as $W$+jets. The value of the $\ttbar$ cross section is scanned until a minimum is found between the expected sum of the backgrounds plus signal and the observed number of data events. The observed cross section is $\sigma_\ttbar$ = 7.2 $\pm$ 0.4 (stat) $\pm$ 0.5 (syst) $\pm$ 0.4 (lumi) pb.

The other CDF analysis in the lepton + jets channel makes use of the difference in kinematic properties between the signal and background events~\cite{topXs_lj_kin_cdf}. Unlike the other analyses presented in these proceedings, that count the number of events above the background to obtain the cross section for the signal, this analysis relies on a fit to a Neural Network (NN) output that is based on inputs from seven kinematic quantities, most of them related to the characteristics of jets in the events. The distribution of the NN output is shown in Fig.~\ref{fig:topXs_lj} (right); it clearly shows the signal peaking at high values and the backgrounds, mostly $W$+jets, peaking at low values. Doing this fit allows us to remove the need for heavy-flavor tagging which has a relatively low efficiency. This analysis has over 1200 reconstructed $\ttbar$ events, and in effect, trades off additional statistics for uncertainties due to the fit of the NN output. This results in a statistical uncertainty comparable between the two lepton + jets measurements. The fitted cross section is $\sigma_\ttbar$= 6.8 $\pm$ 0.4 (stat) $\pm$ 0.6 (syst) $\pm$ 0.4 (lumi) pb.

\subsection{\boldmath{$\ttbar$} Cross Section Combinations}
\begin{figure}
  \begin{center}
     \includegraphics[height=80mm]{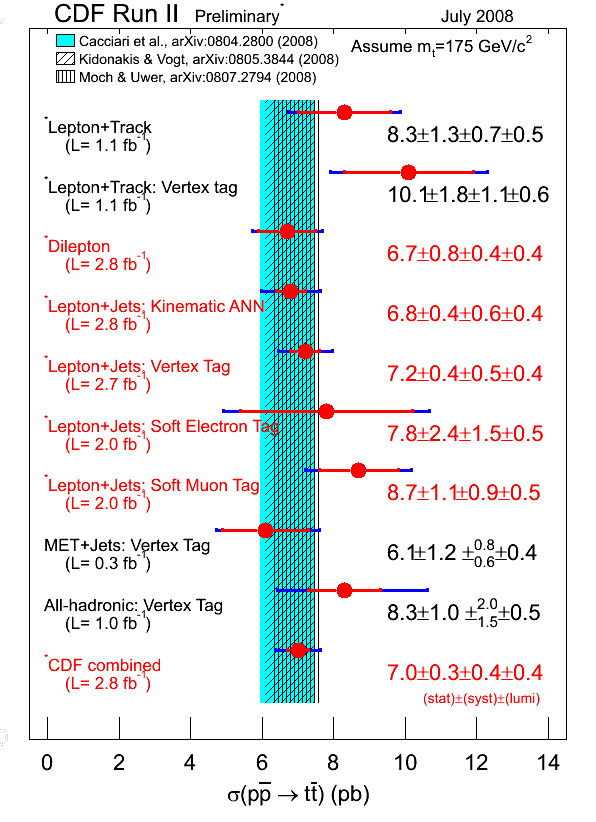}
          \includegraphics[height=80mm]{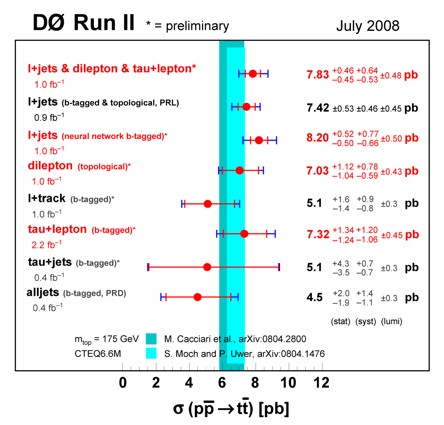}
  \end{center}
  \caption{$\ttbar$ cross section measurements compared to theory (vertical bands): (left) Individual results as well as the combined result from CDF. The measurements in red enter into the combination. (right) Individual and combined results from D\O\. The most recent results are shown in red. For both figures, the red lines show the statistical uncertainty, the blue show the total uncertainties.}
  \label{fig:topXs_combo}
\end{figure}

To obtain the most precise $\ttbar$ cross section, we combine the different measurements. A new combination from CDF was presented at this conference~\cite{topXs_combo_cdf}. The five most recent $\ttbar$ cross section results, shown in Fig~\ref{fig:topXs_combo} (left) are combined. The combined result, also shown on this plot, is $\sigma_\ttbar$ = 7.0 $\pm$ 0.3 (stat) $\pm$ 0.4 (syst) $\pm$ 0.4 (lumi). The total uncertainty is thus of the order of 9\%. The improvements on the single best measurement, the b-tagged lepton + jets measurement described earlier, is of the order of 16\%.

Figure~\ref{fig:topXs_combo} (right) shows a similar plot for the most precise D\O\ measurements, using up to 1.0 $fb^{-1}$ of data~\cite{topXs_combo_d0}. The top line shows a new combination of all analyses, which yields a measured cross section of $\sigma_\ttbar$ = 7.83 $^{+ 0.46}_{ - 0.45}$ (stat) $^{+ 0.64}_{ - 0.53}$ (syst) $\pm$ 0.48 (lumi).

\section{FORWARD-BACKWARD ASYMMETRY IN \boldmath{$\ttbar$} PRODUCTION}
\begin{figure}
  \begin{center}
         \includegraphics[height=65mm]{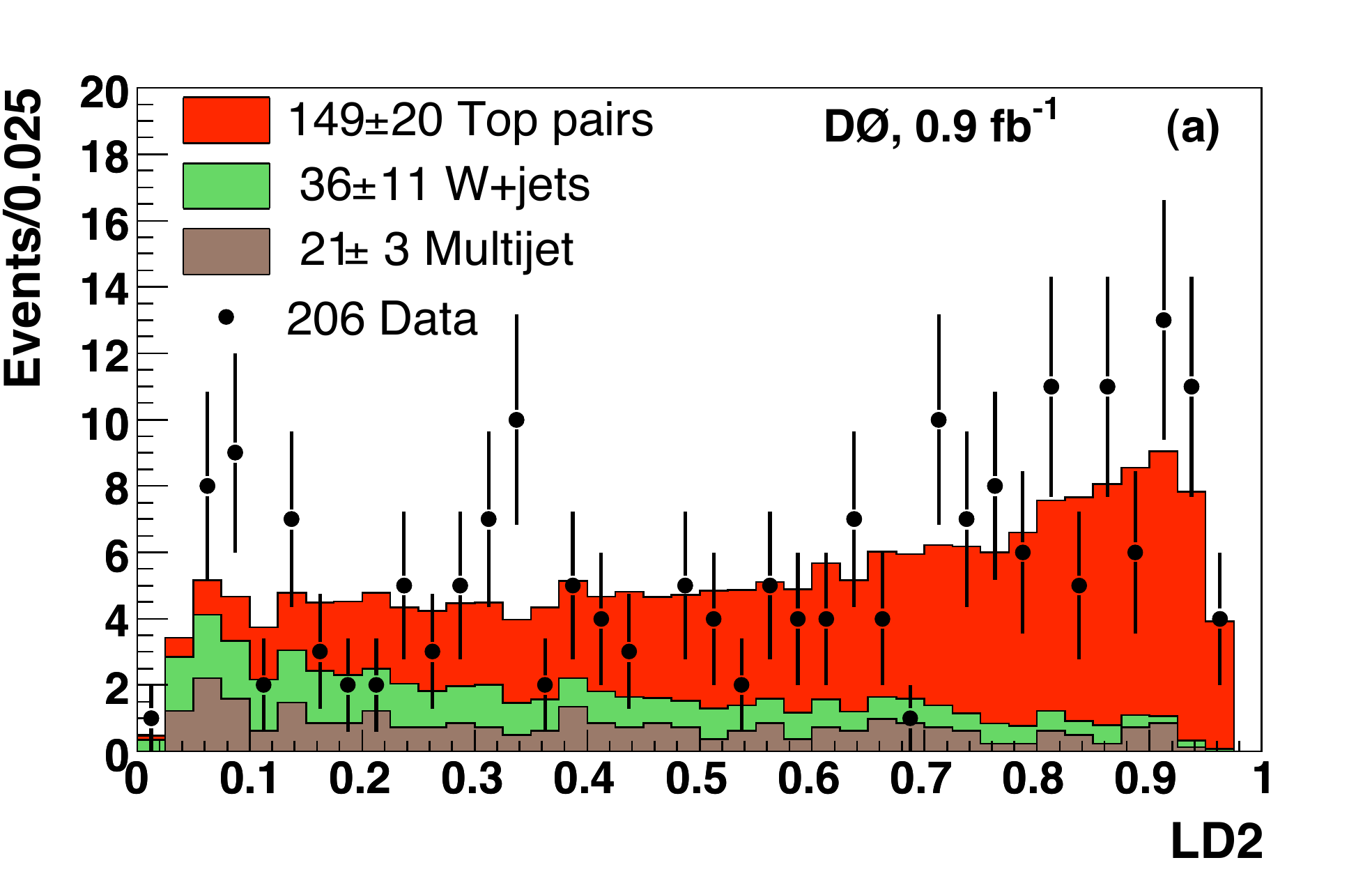}
     \includegraphics[height=65mm]{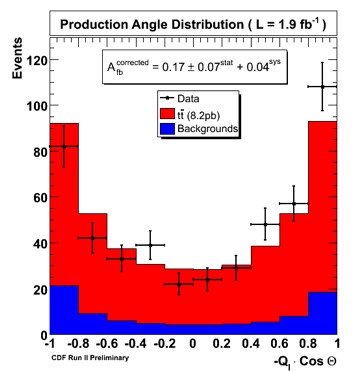}
  \end{center}
  \caption{Forward-backward charge asymmetry of the $t$ and $\bar{t}$: (left) D\O\ : Distribution of the variable that discriminates between top-pair production and background processes for events where the top quark is more forward. The $\ttbar$ signal is shown in red. (right) CDF: $t$ and $\bar{t}$ production angle distribution in the lab frame (with $\cos\theta$ multiplied by the charge of the lepton) showing the data compared to the standard model expectations. The data seem to prefer a larger value of the asymmetry.}
  \label{fig:Afb_cdf}
\end{figure}

The D\O\ Collaboration recently published the first measurement of the forward-backward charge asymmetry in top 
production using 0.9 $fb^{-1}$~\cite{Afb_d0}. While a small asymmetry of the order of 5\% is expected to arise from QCD contributions~\cite{Afb_theory}, other non Standard Model phenomena could drastically affect this asymmetry~\cite{Afb_new_theory}. The asymmetry $A_{fb}$ is defined as the ratio $\frac{N_f - N_b }{N_f +N_b}$, where an event is counted as forward (backward) if the rapidity difference $\Delta y = y_t -  y_{\bar{t}}$ is positive (negative). This measurement is performed in the reconstructed $\ttbar$ rest frame. The result in the inclusive, $\geq$ 4 jets, case is $A_{fb}^\mathrm{uncorr} = 12 \pm 8 \,\mathrm{(stat)}\, \pm 1\, \mathrm{(syst)} $, is not corrected for kinematic acceptance. Figure~\ref{fig:Afb_cdf} (left) shows the discriminant that is used to extract the signal fraction, $N_f$, in the forward region, a similar plot is available for the backward region.
Similar searches have also been performed by the CDF Collaboration using 1.9 $fb{-1}$ of data~\cite{Afb_cdf}. $A_{fb}$ was measured in both the reconstructed rest frame of the $\ttbar$ system and in the laboratory frame, and corrected to the intrinsic parton-level to provide an easy comparison between the experimental $A_{fb}$ result and the theoretical predictions. The results are $A_{fb}^\mathrm{corr} = 24 \pm 13\, \mathrm{(stat)}\, \pm 4 \,\mathrm{(syst)} $and $A_{fb}^\mathrm{corr} = 17 \pm 7 \,\mathrm{(stat)}\, \pm  4 \,\mathrm{(syst)}$, in the $\ttbar$ and lab frame, respectively. Figure~\ref{fig:Afb_cdf} (right) shows the angular distribution of the $t$ and $\bar{t}$ in the laboratory frame, with $\cos\theta$ multiplied by the charge of the lepton, for the CDF analysis performed in the lab frame.
While all measured values of the asymmetry are somewhat larger than the expected \~5\% (which depends on number of jets in the event and their transverse momentum) there is reasonable agreement with the Standard Model.

\begin{thebibliography}{99}
\bibitem{topXs_dil_cdf}
T.~Aaltonen {\em et al.} (CDF Collaboration), CDF Public Note 9399 (2008).

\bibitem{topXs_ltau_d0}
V.~M.~Abazov {\em et al.} (D\O\ Collaboration), D\O\ Note 5607-CONF (2008).
\bibitem{topXs_lj_tag_cdf}
T.~Aaltonen {\em et al.} (CDF Collaboration), CDF Public Note 9462 (2008).
\bibitem{topXs_lj_kin_cdf}
T.~Aaltonen {\em et al.} (CDF Collaboration), CDF Public Note 9474 (2008).
\bibitem{topXs_combo_cdf}
T.~Aaltonen {\em et al.} (CDF Collaboration), CDF Public Note 9448 (2008).

\bibitem{topXs_combo_d0}
V.~M.~Abazov {\em et al.} (D\O\ Collaboration), PRL {\bf 100}, 192004 (2008).

\bibitem{Afb_d0}
V.~M.~Abazov {\em et al.} (D\O\ Collaboration), Phys. Rev. Lett. {\bf 100}, 142002 (2008).
\bibitem{Afb_theory}
 A. Oscar, J. Kuhn, and G. Rodrigo, Phys. Rev. D 77, 014003 (2008); M. T. Bowen, S. Ellis, and D. Rainwater, 
Phys. Rev. D 73, 014008 (2006); S. Dittmaier, P. Uwer, and S. Weinzierl, Phys. Rev. Lett. 98, 262002 (2007); 
D. Almeida, G. Sterman, and W. Vogelsang, arXiv:0805.1885 (2008). 
\bibitem{Afb_new_theory}
 For example, O. Antunano, J. H. Kuhn and G. Rodrigo, arXiv:0709.1652. J. Rosner, Phys. Lett. B 387, 113 
(1996); P. Frampton and S. Glashow, Phys. Lett. B 190, 157 (1987); L. Sehgal and M. Wanninger, Phys. Lett. 
B 200, 211 (1988). 
\bibitem{Afb_cdf}
T.~Aaltonen {\em et al.} (CDF Collaboration),  arXiv:0806.2472, Accepted by PRL (2008).

\end{thebibliography}
\end{document}